# Microfluidic Device for Continuous Magnetophoretic Separation of Red Blood Cells

Ciprian Iliescu[1], Elena Barbarini[3], Marioara Avram[2], Guolin Xu[1] and Andrei Avram[2]

[1] Institute of Bioengineering and Nanotechnology, Singapore

[2] National Institute for Research and Development in Microtechnologies, Bucharest, Romania, Erou Iancu Nicolae Str., 32 B, P.O. Box 38-160, phone: +40214908085, fax: +40214908238

[3] Politecnico di Torino, Electronics Department, 24, Corso Duca degli Abruzzi, 10129 Torino, Italy

*Abstract* - **This paper presents a microfluidic device for magnetophoretic separation red blood cells from blood under continuous flow. The separation method consist of continuous flow of a blood sample (diluted in PBS) through a microfluidic channel which presents on the bottom "dots" of ferromagnetic layer. By applying a magnetic field perpendicular on the flowing direction, the ferromagnetic "dots" generate a gradient of magnetic field which amplifies the magnetic force. As a result, the red blood cells are captured on the bottom of the microfluidic channel while the rest of the blood is collected at the outlet. Experimental results show that an average of 95 % of red blood cells is trapped in the device.**

I. INTRODUCTION

Separation of red blood cells (RBCs) from whole blood is often an essential step before the application of many clinical and molecular diagnostic tests on blood samples. The procedure is also useful in blood transfusion applications as it allows the selective transfusion of particular cell types. Separation can be possible due to hemoglobin. Hemoglobin is a conjugated metal-protein comprising of four polypeptide globins chains containing a ring structure covalently bonded with a central ferrous iron atom ($Fe^{2+}$), which binds reversibly with oxygen. If is deoxygenated, each of the four iron atoms contains four unpaired electrons, giving the protein and the cell a substantial paramagnetic moment. White blood cells (WBCs) do not contain hemoglobin and are diamagnetic particles. These properties are the starting point for the trapping of RBCs particles using high gradient magnetic separation.

Here, we report a simple microfluidic device for extraction of RBCs from blood under continuous flow. The device consists of a microfluidic channel with ferromagnetic "dots" deposited on its bottom. The geometry of the ferromagnetic layer and the application of an external magnetic field perpendicular to the flow direction generate a magnetic force on the RBCs present in the blood. The device offers other advantages, such as the collection of the RBCs inside the channel. The glass structure allows the complete visibility and analysis of the sample.

II. DEVICE DESIGN

The structure of the device is presented in Fig. 1. A glass die with inlet / outlet holes and a 60μm-depth microfluidic channel is bonded to another glass die, on which a ferromagnetic structure – square "dots" (2x2x2μm) of Ni – was patterned. A permanent magnet generates an external magnetic flux of 0.2 T perpendicular on the flow direction. The blood is diluted with PBS and is flown through the microfluidic channel. The ferromagnetic "dots" generate a gradient of magnetic field which amplifies the magnetic force that acts on RBCs. As a result the RBCs are trapped by the ferromagnetic layer while the WBCs are flushed out with the plasma and the other blood components.

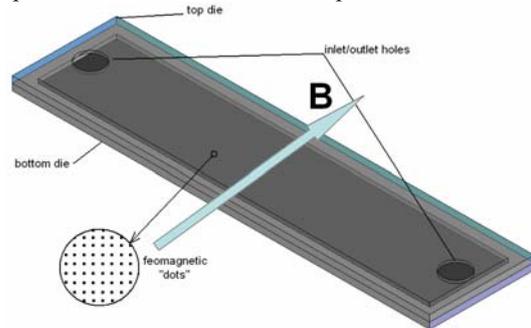

Figure 1: Schematic view of the microfluidic device for RBCs trapping

III. ANALYTICAL CONSIDERATIONS

If a uniform external magnetic field is applied normal to the axis of a ferromagnetic wire, it is deformed near the ferromagnetic wire, and generates a high gradient magnetic field, experienced by the magnetic particles moving around the wire. Therefore, blood cells flowing near the ferromagnetic wire experience a magnetic force created by the high gradient magnetic field near the ferromagnetic wire.

The magnetic force of a rectangular wire on blood cells, located around the x-axis in Figure, can be calculated as

$$F_{BC} = -\frac{2k\mu_0 \Delta\chi V_{BC} a^2}{r^3}\left[k\left(\frac{w}{h}\right)\frac{a^2}{r^2} + \cos 2\varphi\right] \cdot$$

$$\cdot \left(\frac{w}{h}\right)H_0^2 a_r - \frac{2k\mu_0 \Delta\chi V_{BC} a^2}{r^3}\sin 2\varphi \left(\frac{w}{h}\right)H_0^2 a_\varphi, r > a$$

with

$$k = \frac{\mu_w - \mu_B}{\mu_w + \mu_B}$$

where $\Delta\chi$ is the relative magnetic susceptibility of a blood cell relative to the buffer solution; $\mu_w$ and $\mu_B$ are the permeabilities of the ferromagnetic layer and the buffer solution, respectively; VBC is the volume of the blood cell; a is the lateral dimension of the ferromagnetic structure; r and φ are the cylindrical coordinates of the distance and angle; $H_0$ is the external magnetic field; and $a_r$ and $a_φ$ are unit vectors for the distance and angle in the cylindrical coordinates.

If the ferromagnetic structure is magnetically saturated, the first term of the magnetic force in the equation is independent of the external magnetic field, $H_0$, and proportional to the square of the saturation magnetization field. The second and third terms of the magnetic force are linearly proportional to the saturation magnetization field and the external magnetic field.

For magnetic particles placed on the x-axis (φ ≈ 0), sin2φ ≈ 0, cos2φ ≈ 1, the ferromagnetic material attracts paramagnetic particles (RBCs) and repels diamagnetic particles (WBCs). For magnetic particles placed on the y-axis (φ ≈ 90), sin2φ ≈ 0, cos2 φ ≈ -1, the material attracts diamagnetic particles (WBCs) and repels paramagnetic particles (RBCs). The first geometric configuration has been called the paramagnetic capture (PMC) mode; the latter has been called the diamagnetic capture (DMC) mode. To summarize, the main parameters that influence the magnetic force applied on the particles affected by and external magnetic field are: the intensity and orientation of the magnetic field $H_0$; the permeability of the buffer solution and the ferromagnetic material; the susceptibilities of the cells and the buffer solution and the dimension of the device.

## IV. FABRICATION PROCESS

The fabrication process of the device consists of three important steps: fabrication of the top wafer with the inlet/outlet holes and the microfluidic channel, fabrication of the bottom wafer with ferromagnetic concentrators and assembling of the wafers.

The fabrication of the top wafer is presented in Figure 2. A 4" glass wafer Corning 7740 was used. A Cr/Au/ photoresist masking layer (Figure 2a) was used for fabrication of the microfluidic channel. A wet etching process in HF/HCl was used for 60μm deep-etching in glass (Figure 2b). During the wet etching process the backside of the wafer was protected by a dummy silicon wafer bonded with wax on the glass wafer. The debonding of the wafer was performed on a hot plate. After removing of the Cr/Au/photoresist mask (Figure 2c) using classical resist stripper (NMP) – for photoresist and residual wax and Au and Cr etchants, a second Cr/Au/photoresist mask was applied on the other surface of the glass wafer (Figure 2d). Similar, a second wax bonding process will assure the protection of the microfluidic channel, while the wax served also as etch-stop layer for the deep wet etching process in HF/HCl (Figure 2e). Finally, the glass wafer was debonded from the dummy silicon wafer and the masking layer was removed with a similar procedure as was described before.

The fabrication of the bottom wafer consists of a simple lift off process presented in Figure 3. On a 4" glass wafer (Corning 7740) a 5μm-thick photoresist mask (AZ4620 from Clariant) was deposited (Figure 3a). The metal layer Ti/Ni (50nm/2μm) was deposited in a CHA e-beam evaporator – (Figure 3b). The photoresist masking layer was finally removed in an ultrasonic bath using acetone.

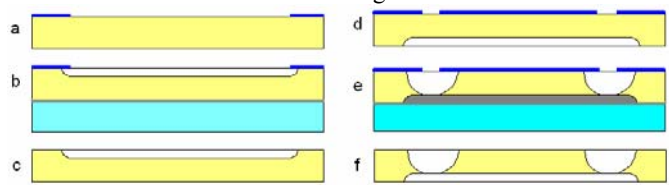

Figure 2: Main steps of the fabrication process of the top wafer: a) processing of the Cr/Au/ photoresist masking layer, b) wet etching of the microfluidic channel, c) removing of the masking layer, d) processing of the Cr/Au/photoresist masking layer for inlet/outlet holes, e) wet etching of inlet/outlet holes, f) removing of the second masking layer.

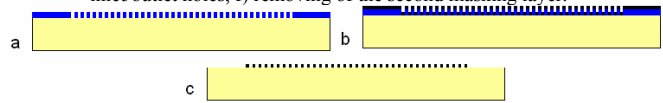

Figure 3: Main steps of the bottom wafer fabrication process: a) deposition of the photoresist mask, b) deposition of Ti/Ni layer, c) removing of the photoresist mask.

For the assembling of the glass wafers a simple adhesive bonding, using SU8-5, photoresist process was used. The technique consists of imprinting of an SU8-5 layer initially spun on a dummy silicon wafer (Figure 4a) on a Teflon cylinder (Figure 4b) and transferring further the adhesive from the cylinder to the top glass wafer (Figure 4c). This contact imprinting technique allowed deposition of a thin adhesive layer only on the bonding regions. In the last step, the wafers are manually align and bonded at 150OC for 30 minutes with an applied pressure of 1000N (Figure 4d). Finally, the wafer is diced. An image of the fabricated device is presented in Figure 5. The dimensions of the chip are 32mm x 9mm.

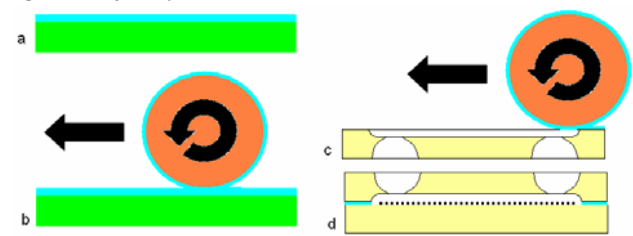

Figure 4: Assembling of the wafers: a) deposition of SU8 on a dummy wafer, b) imprinting of the SU8 on a Teflon cylinder, c) imprinting SU8 on the bonding area, d) alignment and bonding.

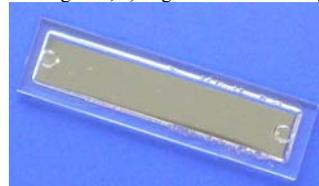

Figure 5. Photo of the microfluidic chip for RBCs trapping

## V. TESTING

Diluted blood (1:20 in PBS) was used for testing purpose. The permanent magnet creates an external magnetic flux of 0.2 T and the flow rate was set in the range between 0.5 and 0.7 ml/h (using a syringe pump). Two connectors fabricated by polymer printing secured the inlet and outlet connections. The results are based on the analysis of the quantity of red cells collected at the output of the device. Experimental results show an average of 5% of red blood cells collected at the output of the device. Figure 6 presents the image with the field densities of cells before and after flowing the sample through the microfluidic device.

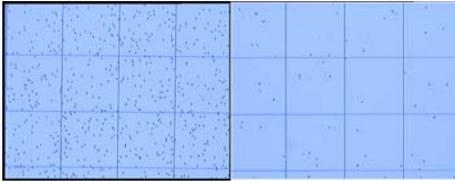

*Figure 6. Image of the field density of the cells on Neubauer hemocytometer before and after flowing the blood sample through the microfluidic device*

## V. CONCLUSIONS

The paper presents a microfluidic device for magnetophoretic trapping of red blood cells under continuous flow, using high gradient of magnetic field. Experimental results show a good trapping efficiency that can be further improved.

REFERENCES

[1] M. Zborowski, G. R. Ostera, Lee R. Moore, S. Milliron, J.J. Chalmers, A.N. Schechtery, Biophysical Journal Vol. 84, 2003 pp. 2638–264.
[2] D. W. Inglis, R. Riehn, R. H. Austin, J. C. Sturm, Applied Physics Letters Vol. 85, No.21, 2004, pp. 5093-5095.
[3] M. Toner, D. Irimia, Annual. Rev. Biomed. Eng, 2005, pp. 77–103.
[4] S. ROATH, A. Richards SMITH, J.H.P. WATSON, Journal of Magnetism and Magnetic Materials 85, 1990, pp.285-289